\documentclass[aps,prb,preprint,showpacs,groupedaddress]{revtex4}
\usepackage[dvips]{graphicx}
\begin{document}
\title{ Sharp magnetization step across the ferromagnetic to antiferromagnetic  transition in doped-CeFe$_2$ alloys}
\author{S. B. Roy} 
\author{M. K. Chattopadhyay}
\author{P. Chaddah}
\affiliation{Low Temperature Physics Laboratory, Centre for Advanced Technology, Indore 452013, India.}
\author{A. K. Nigam}
\affiliation{Tata Institute of Fundamental Research, Mumbai 400005, India.}
\date{\today}
\begin{abstract}
Very sharp magnetization step is observed across the field induced antiferromagnetic to ferromagnetic transition in various doped-CeFe$_2$ alloys, when the measurement is performed below 5K. In the higher temperature regime (T$>$5K) this transition is quite smooth in nature. Comparing with the recently observed similar behaviour in manganties showing colossal magnetoresistance and magnetocaloric material Gd$_5$Ge$_4$ we argue that such magnetization step is a generalized feature of a disorder influenced first order phase transition.
\end{abstract} 
\maketitle

\section{Introduction}
Recent studies of magnetic field induced first order antiferromagnetic (AFM) to ferromagnetic (FM) transition in various manganite compounds showing colossal magnetoresistance (CMR) have revealed ultra sharp magnetization steps when the measurements are performed below 5K \cite{1,2,3,4,5,6,7,8}. Such steps are observed in both single crystal and polycrystalline samples \cite{7}. A catastrophic relief of strain built up during the field induced first order transition between AFM and  FM  phase, has been suggested as a possible cause of such striking behaviour \cite{7}. Very similar magnetization step has also been reported for the magnetocaloric material Gd$_5$Ge$_4$  across the field induced AFM-FM transition\cite{9,10,11}. Although belonging to different classes of materials these two systems have the common features of phase-coexistence and magneto-elastic coupling associated with the AFM-FM transition. To highlight the generality of the observed phenomenon we report here the existence of very sharp magnetization step across the field induced AFM-FM transition in Ru and Re-doped CeFe$_2$ alloys belonging to an entirely different class of materials. We argue that such magnetization step is a characteristic feature of  disorder influenced first order magneto-structural phase transition.

CeFe$_2$ is a ferromagnet with Curie temperature T$_C \approx$ 230K\cite{12}.  A small substitution (3-6$\%$) of selected elements such as Co, Al, Ru, Ir, Os and Re  induces a low temperature AFM  state in this otherwise FM compound\cite{13,14}.  The ferromagnetic FM-AFM transition in these alloys is accompanied by a structural distortion and  a discontinuous change of the unit cell volume\cite{15}. Inside the AFM state an application of external magnetic field (H)  induces back the original FM state, while at the same time erases the structural distortion and recovers the original cubic structure. The first order nature of this AFM-FM transition has been emphasised with various kinds of measurements\cite{16,17,18}.  

\section{Experimental}
We use a 4$\%$Ru and a 5$\%$Re doped CeFe$_2$ sample for our present study. The samples were prepared by argon-arc melting starting from metals of at least nominal 99.99\% purity. These polycrystalline samples were characterized with metallography, X-ray diffraction(XRD) and neutron scattering studies \cite{13,14,15}.  Due to the peritectic reaction during the solidification process, one expects to find in the as-cast structure cores of Ce$_2$Fe$_{17}$ with perhaps some iron-solid solution at the centre, surrounded by shells of CeFe$_2$ and the eutectic material. Normally, with adequate heat treatment the first formed solid should disappear. However, in practice there is almost always some trace of second phase in the annealed samples. Indeed the traces of impurity phases were still found after annealing the present samples at 600$^o$C for seven days. With various heat treatments it was found that the sequence of annealing at 600$^o$C for two days, 700$^o$C for 5 days, 800$^o$C for 2 days and 850$^o$C for one day improved the quality of the samples a great deal \cite{13,14}. Combination of metallography\cite{14}, XRD\cite{14} and neutron scattering study\cite{15} indicates that the amount of second phase in these samples is less than 2\%. CeFe$_2$ forms in cubic Laves phase structure. In this structure all of the Ce sites and the Fe sites are equivalent. The Fe atoms form a three-dimensional open network of corner sharing tetrahedra interpenetrated by a diamond structure of Ce atoms. Lattice constant of CeFe$_2$ is $\approx$ 7.3 A$^o$ and it increases with Ru and Re substitutions of Fe.  Magnetization measurements have been performed using a SQUID magnetometer (MPMS5, Qunatum Design) and a vibrating sample magnetometer (VSM, Oxford Instruments).

\section{Results and Discussion} 
In Fig.1 we present the M vs T plots for 4$\%$Ru and 5$\%$Re doped CeFe$_2$ samples. The sharp rise and fall as a function of decreasing temperature indicates the onset of the paramagnetic (PM)- FM  and  FM-AFM  transitions respectively. These results are already known\cite{18,19}, but reproduced here to make the present work self contained.  
\begin{figure}[t]
\centering
\includegraphics[width = 8 cm]{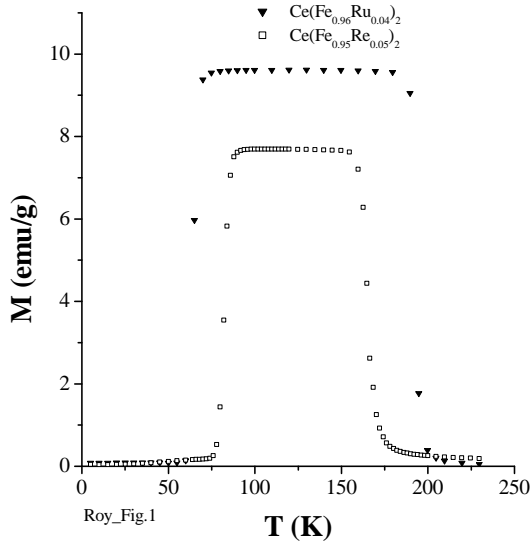}
\caption{Magnetization versus temperature plots for Ce(Fe$_{0.96}$Ru$_{0.04}$)$_2$ and Ce(Fe$_{0.95}$Re$_{0.05}$)$_2$ obtained in the zero field cooled mode in an applied field of 100 Oe.The data presented in this figure are taken from Ref. 18 and 19 respectively}
\label{fig1}
\end{figure}

In fig. 2 ( 3 ) we present isothermal M-H curves at 2.2 K (3K) for 4$\%$Ru (5$\%$Re)  doped CeFe$_2$ samples. On increasing H from zero after zero field cooling (ZFC) from above 200K there is a monotonic rise in M for  4$\%$ Ru (5$\%$ Re) sample below $\approx$45 kOe (69 kOe). Actually there is a finite non-linearity in the very low field regime which is more visible in the 4$\%$Ru-doped sample. This has been attributed to the intrinsic canted character of the AFM state\cite{20,21}. The onset of  the AFM-FM transition is identified with the distinct rapid increase in M around 45 kOe (69 kOe) in the 4$\%$Ru (5$\%$Re)  sample. This behaviour is highlighted for the 4$\%$Ru doped sample in the bottom inset of Fig. 2. This initial rapid increase in M is followed by a very sharp step  in the isothermal M-H curve at 48 kOe(73 kOe) for the  4$\%$Ru (5$\%$ Re). The sharpness of this step is further highlighted in the bottom inset of Fig.3. In the decreasing H cycle the reverse transformation from the FM state to the AFM state is relatively more gradual and the onset of the magnetization step is also delayed. The difference in the nature of the transition process in the increasing H and decreasing H cycle in the doped CeFe$_2$ samples has earlier been attributed to the asymmetry between supercooling and superheating across a first order phase transition\cite{17}. The measurement in that work, however, was confined to the temperature regime T$\geq$5K and the step in the magnetization was not visible. Such asymmetry of the transition between the increasing and decreasing H cycle has not been reported for the CMR-manganites and Gd$_5$Ge$_4$. In the latter compound of course  the AFM-FM transition is irreversible below 20K, namely the FM state is not reverted back to the AFM state on reducing H\cite{10}.  In the present measurement after reducing H to zero we reversed the direction of the applied field and the magnetization step occurred in the same H values in the increasing and decreasing H cycle (see Fig.2 and 3.). After completing this M-H cycle we raised H in the original direction. The magnetization step was retraced as in the increasing H cycle in the third quadrant (in Fig.2 and 3) but the data are not shown here for the sake of clarity of the initial ZFC virgin curve. All these studies clearly show the robustness of this sharp magnetization step. 
\begin{figure}[t]
\centering
\includegraphics[width = 8 cm]{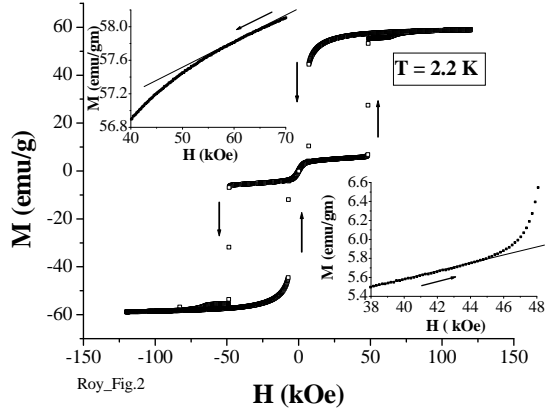}
\caption{Isothermal variation of magnetization of Ce(Fe$_{0.96}$Ru$_{0.04}$)$_2$ 
sample with applied magnetic field at T=2.2 K. The field is increased to 120 kOe with a rate of 1 kOe/min starting from the zero field cooled state and then reduced to zero in the same rate. The field direction is then changed and the same cycle is repeated with the same rate. The bottom (top) inset highlights the onset of the AFM-FM (FM-AFM) transition in the increasing (decreasing) field cycle in the positive field direction.}
\label{fig2}
\end{figure}

We have shown earlier that there is a finite field sweep rate dependence of magnetization in the AFM-FM transition region of the doped-CeFe$_2$ alloys. However, this sweep rate dependence did not lead to any qualitative change in magnetic behaviour between the increasing and decreasing H cycle. In the present study we have used a relatively slow sweep rate of 1 kOe/min while ensuring the temperature stability during the experimental time ($\approx$ 10 hrs) of the isothermal M-H measurements in the temperature regime below 5K. With this field sweep rate the magnetization step was earlier observed both in Gd$_5$Ge$_4$ and manganites (see Ref.11).   

\begin{figure}[t]
\centering
\includegraphics[width = 8 cm]{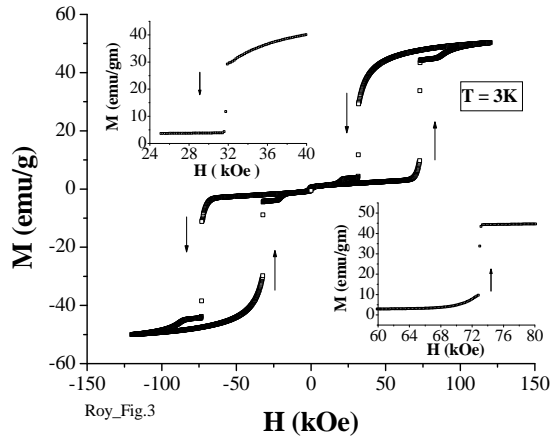}
\caption{Isothermal variation of magnetization of Ce(Fe$_{0.95}$Re$_{0.05}$)$_2$ 
sample with applied magnetic field at T=3K. The field cycle and the field ramp rate is the same as in Fig.2. The bottom (top) inset highlights the large magnetization step across the AFM-FM (FM-AFM) transition in the field increasing (decreasing) cycle in the positive field direction.}
\label{fig3}
\end{figure}

In Fig. 4(a) and (b) we show  isothermal M-H curves at 10K for 5$\%$Re and 4$\%$Ru doped CeFe$_2$ samples respectively. Although the onset of the AFM-FM transition is clearly marked by the sharp increase in M, there is no sign of any sharp step in the M-H curve. On the other hand the asymmetry\cite{17} in the transition between the increasing and decreasing H cycles is clearly visible in both the samples.
\begin{figure}[t]
\centering
\includegraphics[width = 8 cm]{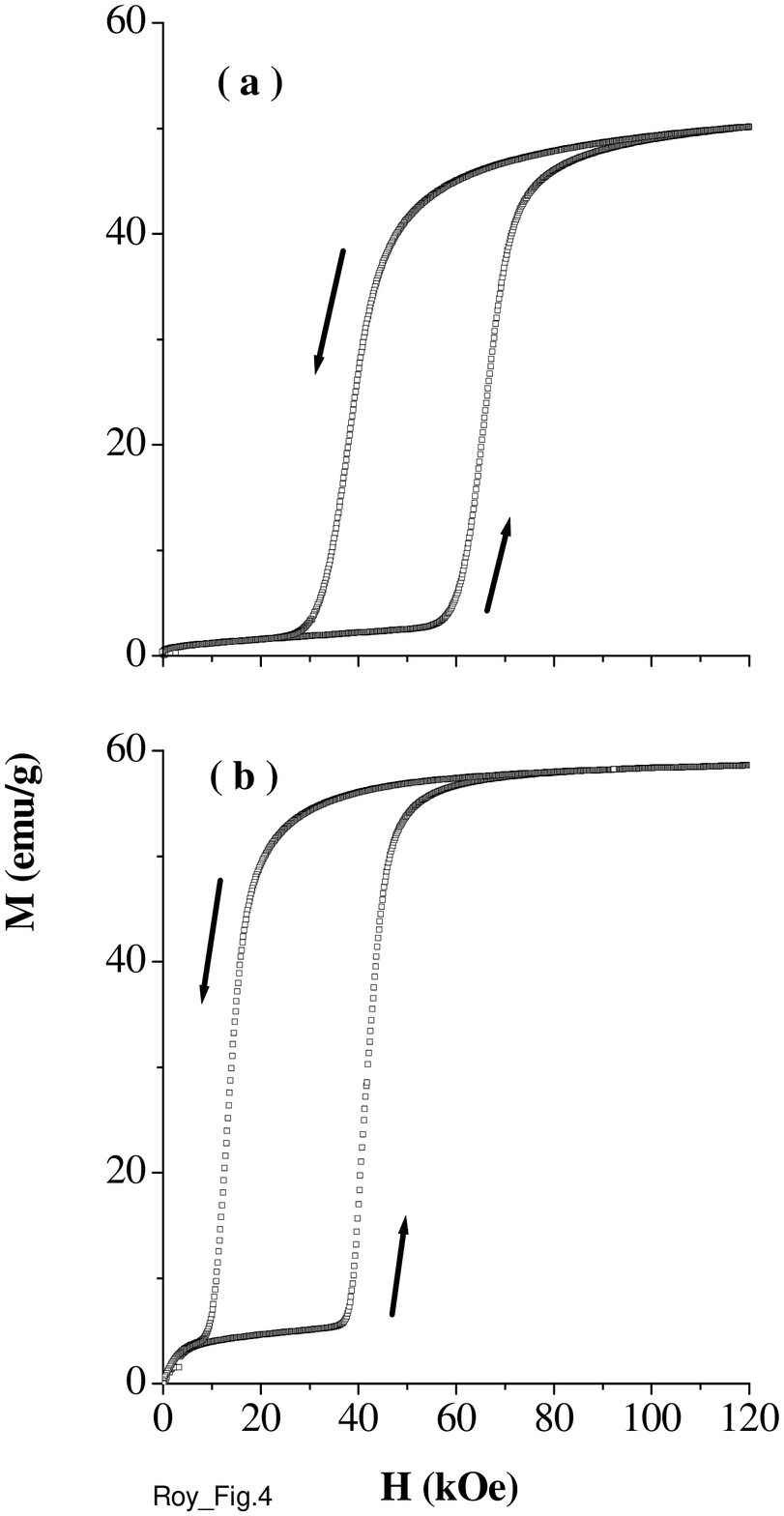}
\caption{Isothermal variation of magnetization of (a) Ce(Fe$_{0.95}$Re$_{0.05}$)$_2$ (b) Ce(Fe$_{0.96}$Ru$_{0.04}$)$_2$ 
samples with applied magnetic field at T= 10K. The field cycle and the field ramp rate is the same as in Fig.2 and 3. The M-H loop in the negative field direction is not shown here for the sake of conciseness. Note that the large magnetization steps seen in Fig.2 and 3 are not visible here.}
\label{fig4}
\end{figure}

It is to be noted that the onset of the FM state in the increasing H cycle  occurs at a lower field than does nucleation of the AFM phase during  the decreasing H cycle. This is highlighted in the insets of fig.2 for the 4$\%$Ru-doped sample. The onset of the transition is around 45 kOe in the increasing H cycle (lower inset), while it is definitely above 55 kOe in the decreasing H cycle (upper inset). Similar results exist for the 5$\%$Re-doped sample but not shown here for the sake of conciseness. This behaviour was argued earlier as an indication of local variation of the AFM-FM transition temperature/field in the sample\cite{17,18}. The composition in any alloy or doped compound is a spatially inhomogeneous quantity, and the actual composition will vary around some average composition simply due to the disorder that is frozen in as the solid crystallizes from the melt\cite{22}. Since the transition temperature is  sensitive to composition, there must exist a locally defined hypothetical transformation temperature, which depends on local composition. Such distribution or landscape of transition field gives rise to the impression of global rounding of the transition in the bulk magnetization measurements. This idea is in consonance with the disorder influenced first order transition proposed by Imry and Wortis\cite{23}. A very similar disorder induced rough landscape picture has  been proposed for the vortex solid melting in the high T$_C$ superconducting material BSCCO \cite{24}. The applicability of this landscape picture in doped-CeFe$_2$ alloys has now been confirmed with micro-Hall probe imaging \cite{25}. However, there remained some question regarding the relevance of theoretical models with uncorrelated  disorder in explaining  the observed  size of  AFM/FM phase clustering in  several micrometer scale \cite{25}. Very recently the effect of strain-disorder coupling across such disorder influenced first order transition has been studied theoretically and phase-coexistence in the micrometer scale has actually been predicted \cite{26,27}. While some evidence of magneto-elastic coupling already existed in the earlier studies on doped-CeFe$_2$ alloys \cite{15,28}, recent magnetostriction measurement\cite{29} has further established the role of magnetoelastic coupling on the first order AFM-FM transition. At this stage it becomes natural to put forward the argument that the magnetization steps are linked to the catastrophic relief of strain build up during the first order magneto-structural transition. However, the absence of the steps in the isothermal magnetization measurements in higher T regime still needs to be explained. 

We shall now attempt to understand the observed behaviour at all temperature regimes within a framework of local distribution of transition temperature/field. In the absence of any disorder mediated heterogeneous nucleation a system reaches a metastability limit \cite{30} well beyond the thermodynamic transition point before a jump takes place from one phase to the other. In a rough landscape picture such jumps give rise to a series of steps in the measurable quantities like magnetization and specific heat. However, in the presence of active nucleation centres these steps are replaced by a continuous change giving the impression of  a broadened transition. The fluctuational development of nuclei in a size range around  a critical size determined by the material parameters is an essential part of the kinetics of  first order phase transition\cite{31}. The distribution function for nuclei of  various sizes actually broadens with the increase in temperature \cite{31}. This picture can explain the smooth behaviour across the metamagnetic transition observed in our present samples above 5K (see Fig.4). In the T regime below 5K the intrinsic thermal energy fluctuation arising from the k$_B$T term becomes quite small making many of the nucleation barriers insurmountable, hence effectively reducing the number of nucleation centres. This will increase the possibility of the step like features in various observables in this lower temperature regime. It is worth recalling here that the quenched disorder in our present system mainly consists of purely statistical compositional disorder\cite{22}. Hence in the proposed landscape picture the distribution of  transition temperature/field will have peak at the target compositional  value of the sample with tails on either side. This will lead to a big step in global measurements of magnetization with smaller steps on either side. In addition in systems like doped-CeFe$_2$ alloys with appreciable coupling between electronic and elastic degrees of freedom, an applied magnetic field can lead to magneto-elastic coupling between different regions in the sample. This is also likely to encourage a single big step in the field dependence in M across a magneto-structural transition. While we do see this big step in magnetization below 5K, we are unable to resolve the smaller steps in our polycrystalline sample. We believe these smaller steps can be observed in a single crystal sample with a landscape of less roughness and going further down in temperature. It is to be mentioned here that the measurement procedure to obtain the results in Fig.2 and 3 did not engage any heater in the sample chamber. A heater was operational in the present study for active temperature control in the temperature regime T$\geq$5K.  Thus, considering how the temperature of the exchange gas is controlled by a temperature controller (i.e., a heater and a feedback loop), and taking into account that critical magnetic field for the onset of AFM-FM transition in CeFe$_2$ alloys is strongly dependent on temperature one can not rule out the possibility that  temperature fluctuations on the order of 0.1 K  trigger a transformation of a large fraction of a material even when H is held constant. Hence this extrinsic source of temperature fluctuations is likely to add to the intrinsic thermal fluctuations (i.e. k$_B$T term) in smoothening out the M-H curve across  the AFM-FM transition in the higher T regime.

There exist some theoretical studies of field driven first order transition based on both random bond \cite{32} and random field Ising models \cite{32,33} with quenched disorder. With varying amount of disorder the nature of the non-equilibrium transition changes from a discontinuous one with one or more large avalanches to a smooth one with only tiny avalanches. Projecting to our present experimental studies it can be argued that at higher T most of the available quenched disorder sites remain active. At lower T the k$_B$T term is smaller than the local nucleation barriers at many of the quenched disorder site. This renders such disorder sites ineffective for nucleation. We would also like to mention here that the discrete steps in the magnetization observed across the AFM-FM transition in good quality polycrystalline samples of Gd$_5$Ge$_4$, disappears in more disordered samples giving rise to a smooth change\cite{34}.

\section{Conclusion}
In conclusion we have observed a very sharp magnetization step across the field induced AFM-FM transition in Ru and Re doped-CeFe$_2$ alloys, when the measurement is performed in the temperature regime below 5K. We have tried to understand this interesting feature within the frame work of a disorder-influenced first order magneto-structural phase transition. The observed magnetization step is markedly similar to the step observed across the field-induced AFM-FM transition in various CMR-manganite systems and magnetocaloric material Gd$_5$Ge$_4$. It is now well known that a structural transition accompanies the first order AFM-FM transition in these classes of materials. We have earlier highlighted that the phase-coexistence and metastability are common features across the field/temperature induced AFM-FM transition in CMR-manganites, Gd$_5$Ge$_4$ and doped-CeFe$_2$ alloys, and argued that those arise due to the influence of disorder on a first order magneto-structural phase transition\cite{25,34}.  Combining our present experimental results on doped-CeFe$_2$ alloys with the existing results on various CMR-manganite systems and magnetocaloric material Gd$_5$Ge$_4$, we speculate that the observed sharp step in magnetization across the AFM-FM transition in all these different classes of materials is a universal feature of a disorder influenced first order magneto-structural phase transition.

\end{document}